\documentclass[12pt]{iopart}

\usepackage{iopams}  
\usepackage{bm}
\usepackage{here}
\usepackage{tikz}
\usepackage{circuitikz}
\makeatletter
 \pgfcircdeclarebipolescaled{instruments}
{
    \anchor{text}{\pgfextracty{\pgf@circ@res@up}{\northeast}
        \pgfpoint{-.5\wd\pgfnodeparttextbox}{
            \dimexpr.5\dp\pgfnodeparttextbox+.5\ht\pgfnodeparttextbox+\pgf@circ@res@up\relax
        }
    }
}
{\ctikzvalof{bipoles/oscope/height}}
{josephson}
{\ctikzvalof{bipoles/oscope/height}}
{\ctikzvalof{bipoles/oscope/width}}
{
    \pgf@circ@setlinewidth{bipoles}{\pgfstartlinewidth}
    \pgfextracty{\pgf@circ@res@up}{\northeast}
    \pgfextractx{\pgf@circ@res@right}{\northeast}
    \pgfextractx{\pgf@circ@res@left}{\southwest}
    \pgfextracty{\pgf@circ@res@down}{\southwest}
    \pgfmathsetlength{\pgf@circ@res@step}{0.25*\pgf@circ@res@up}
    \pgfscope
        \pgfpathrectanglecorners{\pgfpoint{\pgf@circ@res@left}{\pgf@circ@res@down}}{\pgfpoint{\pgf@circ@res@right}{\pgf@circ@res@up}}
        \pgf@circ@draworfill
    \endpgfscope
    \pgfscope
      \pgfpathmoveto{\pgfpoint{\pgf@circ@res@left}{\pgf@circ@res@up}}%
      \pgfpathlineto{\pgfpoint{\pgf@circ@res@right}{\pgf@circ@res@down}}%
      \pgfpathmoveto{\pgfpoint{\pgf@circ@res@right}{\pgf@circ@res@up}}%
      \pgfpathlineto{\pgfpoint{\pgf@circ@res@left}{\pgf@circ@res@down}}%
      \pgfusepath{draw}
    \endpgfscope
}
\def\pgf@circ@josephson@path#1{\pgf@circ@bipole@path{josephson}{#1}}
\tikzset{josephson/.style = {\circuitikzbasekey, /tikz/to path=\pgf@circ@josephson@path, l=#1}}

\begin{document}

\title{Gauge invariant quantization for circuits including Josephson junctions}

\author{Hiroyasu Koizumi}

\address{Division of Quantum Condensed Matter Physics, Center for Computational Sciences, University of Tsukuba, Tsukuba, Ibaraki 305-8577, Japan}
\ead{koizumi.hiroyasu.fn@u.tsukuba.ac.jp}
\vspace{10pt}
\begin{indented}
\item[]Feb  2024
\end{indented}

\begin{abstract}
Recently, a new theory of superconductivity has been put forward that
attributes the origin of superconductivity to the appearance of a non-trivial Berry connection from many-electron wave functions.
  This theory reproduces the major results of the BCS theory
with conserving the particle number, and predicts the single-electron supercurrent tunneling across the Josephson junction with keeping the correct Josephson relation.
We re-examine the quantization of superconducting qubit circuits by taking into account the above development, and show that the dynamical variables used in the standard theory, the flux nodes relating to the voltage, should be replaced by those relating to the electromagnetic vector potential.  
The fact that the Josephson junction tunneling allows the single-electron supercurrent tunneling is the reason for the existence of excited single electrons in superconducting qubits with Josephson junctions. We predict that
it will be avoided by weakening the coupling between two superconductors in the
Josephson junction.\end{abstract}

%
%
%
%
%

\section{Introduction}

The transmon type superconducting qubit is the most successful one for building quantum computers at present.
The key element for it is the Josephson (superconductor-insulator-superconductor (SIS)) junction,
and the precision control of it is an important issue. The current major obstacle preventing its high performance is the presence of abundant excited single electrons that cause decoherence of qubit states \cite{poisoning2023,Serniak2019}. This problem is often called, the `quasiparticle poisoning problem'
since it is believed that such excited single electrons are created  through the Bogoliuvos quasiparticle excitations.

The Josephson junction is a junction devised by Josephson to discuss the current flow in superconductors. He predicted peculiar current flow through this junction by the electron-pair tunneling \cite{Josephson62}. 
The derived formula for supercurrent through the junction is given by
\begin{eqnarray}
J=J_c \sin \phi
\label{eqJ1}
\end{eqnarray}
where $J_c$ is the constant depending on the junction, and $\phi$ is the difference of the superconducting phase variable between the two superconductors across the insulator part. The phase $\phi$ appears due to the fact that the BCS superconducting state is a coherent superposition of states with different number of pairing electrons (or `Cooper pairs')\cite{BCS1957}. 
Josephson predicted that if $\phi=\phi_0 \neq \pi n$ ($n$ is a integer),
the supercurrent $J=J_c \sin \phi_0$ should flow; and further predicted that if the voltage across the junction $V$ exists,
$\phi$ should show the time-dependence 
\begin{eqnarray}
{d \over {dt}}\phi={{2eV} \over \hbar}
\label{JJdt}
\end{eqnarray}
where $-e$ is the electron charge, and $\hbar$ is the reduced Plank constant. 
This time-dependence is explained by the fact that $\phi$ is related to the chemical potential difference across the junction that gives rise to the electron-pair transfer energy $2eV$. 
The above predictions were experimentally confirmed, and it is now widely-believed that it is the evidence that supercurrent carriers are electron-pairs.

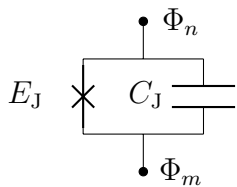
\begin{figure}[H]
\centering
	\begin{circuitikz} \draw
		
		
		(9,0) to [short,*-] (9,0.5)
		
		(8.2,0.5) to (9.8,0.5)
		(8.2,0.5) to[barrier, l=$E_{\rm J}$,] (8.2,1.5)
		(9.8,0.5)  to[C, l=$C_{\rm J}$,] (9.8,1.5)
		(8.2,1.5) to (9.8,1.5)
		(9,1.5) to [short,-*] (9,2)
		
		(9.5,0) node[]{$\Phi_m$}
		(9.5,2) node[]{$\Phi_n$}
		;
	\end{circuitikz}
	\caption{A Josephson junction with capacitance $C_{\rm J}$ and Josephson energy $E_{\rm J}$.}
	\label{JJCircuitElements}
\end{figure}

However, the original Josephson's derivation has two defects, and it is shown that the supercurrent through the Josephson junction that explains the relation in Eq.~(\ref{JJdt}) is actually single-electron tunneling \cite{HKoizumi2015,koizumi2022b}. 
One of the defects is its missing of a capacitance contribution; a Josephson junction is expressed using circuit elements as in Fig.~\ref{JJCircuitElements}; it contains a capacitor part, but this contribution was not included in the original derivation. This missing is caused by the direct application of
the superconductor-normal metal tunneling formalism \cite{Cohen1962} to the Josephson junction case. 

The other defect is that $\phi$ is not gauge invariant since the $U(1)$ phase of the wave function is gauge-dependent,
thus, the current $J$ is actually indeterminate in the original derivation. 
The gauge invariant phase difference for the electron-pair tunneling
is
\begin{eqnarray}
\phi^{\rm old}= \int^{n}_{m}
{{2e} \over \hbar} \left[ {\bf A}({\bf r}, t) -{{\hbar  } \over {2e}} \nabla \chi({\bf r}, t) \right] \cdot d {\bf r}
\label{eq3}
\end{eqnarray}
where $n$ and $m$ denotes the two coordinate points in superconducting electrodes of the Josephson junction, the integration is along the path across the insulator part, ${\bf A}$ is the electromagnetic vector potential, and $\chi$ is an angular variable with period $2\pi$. The phase $-\chi$ corresponds to $\phi$ in the original derivation, where the vector potential ${\bf A}$ is not included.

We derive the time derivative of $\phi^{\rm old}$ in the following for the convenience of the later arguments.
 Let us adopt the gauge in which the electric field ${\bf E}$ is given by ${\bf E}=-\partial_t {\bf A}$, and calculate the time-derivative of $\phi^{\rm old}$. 
The contribution from the first term of the right-hand side of Eq.~(\ref{eq3}) is calculated as
\begin{eqnarray}
\int^{n}_{m}
{{2e} \over \hbar} \partial_t {\bf A}({\bf r}, t) \cdot d {\bf r}=-{{2e} \over \hbar}\int^{n}_{m}
{\bf E}({\bf r}, t) \cdot d {\bf r}={{2e} \over \hbar}V
\label{eq3-A}
\end{eqnarray}
where ${\bf E}$ is the electric field in the capacitor (this contribution is absent in the original derivation).
The contribution from the second term arises from the chemical potential difference (this is the part included in the original derivation), which is calculated as
\begin{eqnarray}
-\partial_t\int^{n}_{m}
 \nabla \chi({\bf r}, t) \cdot d {\bf r}=\partial_t\chi_m-\partial_t\chi_n={ 2 \over \hbar}(\mu_m-\mu_n)={{2e V} \over \hbar}
\label{eq3-Chi}
\end{eqnarray}
where $\mu_n$ denotes the chemical potential at the $n$th node, and the balance between the voltage from the electric field and chemical potential,
$
V=- \int^{n}_{m}
{\bf E}({\bf r}, t) \cdot d {\bf r} ={1 \over {-e}} (\mu_n -\mu_m)
$, is used. Here, the relation $\partial_t\chi_m-\partial_t\chi_n={ 2 \over \hbar}(\mu_m-\mu_n)$ is used; this relation will be found in our previous work \cite{koizumi2022b}.
Due to the presence of the two contributions, the time derivative of $\phi^{\rm old}$ becomes
\begin{eqnarray}
{d \over {dt}}{\phi}^{\rm old}={{4eV} \over \hbar}
\label{eqJosephsonOld}
\end{eqnarray}
which disagrees with Eq.~(\ref{JJdt}).
Actually, the correct one is obtained if we consider single-electron tunneling, and use $\phi^{\rm new}$ given by
\begin{eqnarray}
\phi^{\rm new}={1 \over 2} \phi^{\rm old}=\int^{n}_{m}
{{e} \over \hbar} \left[ {\bf A}({\bf r}, t) -{{\hbar  } \over {2e}} \nabla \chi({\bf r}, t) \right] \cdot d {\bf r}
\label{eq3-new}
\end{eqnarray}
In the standard theory based on the BCS one, the correct Josephson relation in Eq.~(\ref{JJdt}) was obtained using $\phi^{\rm old}$ with neglecting ${\bf A}$. But it is sensible to regard $\phi^{\rm new}$ with including ${\bf A}$ as the correct one.

Recently, a new superconductivity theory is put forward \cite{koizumi2022,koizumi2022b}. In this theory $\phi^{\rm new}$ is obtained with attributing
$\chi$ to the neglected $U(1)$ phase in the Schr\"{o}dinger representation of quantum mechanics by Dirac \cite{Dirac,Dirac1929,Pople}.
The new theory reproduces the major results of the BCS theory \cite{koizumi2022}, resolves some discrepancies that the BCS theory has with experiments \cite{koizumi2022,koizumi2022b,koizumi2023}.
The present work is based on this new theory. In particular we will concern with problems associated with superconducting qubits with Josephson junctions.

Experiments on superconducting qubits exhibit the quasiparticle poisoning problem. They indicate it occurs through single-particle Josephson junction tunneling \cite{poisoning2023,Serniak2019}. Further, the obtained ratio of poisoning single particles to Cooper pairs disagrees with that obtained by the BCS theory; while the former ranges from $10^{-9}$ to $10^{-5}$, the latter  is $10^{-52}$ \cite{Serniak2019}. This huge difference signals something may be fundamentally wrong.
It may be the evidence that the electron  
tunneling across the Josephson junction is single-electron one in an intrinsic manner, and $\phi^{\rm new}$ is the right one. 
 In the present work, we examine this possibility.
 In the due course, we reformulate the quantization of circuits relevant to superconducting qubits based on the new theory. This will require the change of dynamical variables for the circuit quantization.
We will also propose a way to avoid the quasiparticle poisoning. 

The organization of the present work is following: In Section~\ref{sec2}, we discuss the quantization of electronic circuits. We argue that the change dynamical variables is necessary from those using the node voltages to those using the vector potential. 
In Section~\ref{sec3}, we explain how the single electron supercurrent tunneling across Josephson junction is possible.
In Section~\ref{sec4}, a way to avoid the `quasiparticle poisoning' is suggested; it will be achieved by weakening the coupling between two superconductors so that the situation is realized where the $\phi$ in the Josephson relation becomes $\phi^{\rm old}$; in this situation, the relation in Eq.~(\ref{eqJosephsonOld}) will be observed. Lastly, we conclude the present work in Section~\ref{sec5}.

\section{Change of dynamical variables for the quantization of circuits}
\label{sec2}

In this section, 
we consider how ${\phi}^{\rm new}$ is implemented in the quantization of circuit with Josephson junctions. We will change the dynamical variables from the current standard dynamical variables using the volatages at nodes (given below in Eq.~(\ref{eq1}) and adopted in Ref.~\cite{koizumi2022b}) to those using the vector potential (given below in Eq.~(\ref{Phi-A})).

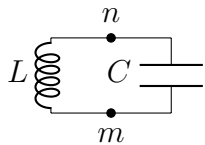
\begin{figure}[H]
\centering
	\begin{circuitikz} \draw
		
		
		 (9,0.5) to [short,*-] (9,0.5)
		
		(8.2,0.5) to (9.8,0.5)
		(8.2,0.5) to[L, l=$L$,] (8.2,1.5)
		(9.8,0.5)  to[C, l=$C$,] (9.8,1.5)
		(8.2,1.5) to (9.8,1.5)
		(9,1.5) to [short,-*] (9,1.5)
		
		(9.,0.2) node[]{$m$}
		 (9.,1.8) node[]{$n$}

		;
	\end{circuitikz}
	\caption{A $LC$ circuit composed of an inductor with inductance $L$ and a capacitor with capacitance $C$.}
	\label{LCCircuitElements}
\end{figure}

In the standard treatment, node fluxes are taken as dynamical variables, where the flux for $n$th node is defined by
\begin{eqnarray}
\Phi_n =  \int^t_{-\infty} V_n(t') dt'
\label{eq1}
\end{eqnarray}
with $V_n(t)$ being the voltage of the $n$th node at time $t$ \cite{PhysRevA.29.1419,Devoret1997,GU20171,RevModPhys.93.025005}.
The voltage between the $n$th and $m$th nodes is given by $\dot{\Phi}_n -\dot{\Phi}_m$, 
and the magnetic flux through the inductor between them is given by ${\Phi}_n -{\Phi}_m$. 
The Lagrangian for the LC circuit depicted in Fig.~\ref{LCCircuitElements}, composed of a capacitor with capacitance $C$ and an inductor with inductance $L$ between the nodes $n$ and $m$, is given by
\begin{eqnarray}
{\cal L}_{LC}={ 1 \over {2}}C \left(\dot{\Phi}_n-\dot{\Phi}_m \right)^2-{ 1 \over {2L}}\left( {\Phi}_n-{\Phi}_m\right)^2
\label{eqLC}
\end{eqnarray}
where the first term corresponds to the energy stored in the capacitor that plays a role of the kinetic energy,
and the second term corresponds to the minus of the energy stored in the inductor that plays a role of the potential energy.
The canonical conjugate variable for ${\Phi}_n$, $Q_n$, is calculated as 
\begin{eqnarray}
Q_n={ {\partial {\cal L}_{LC}} \over {\partial \dot{\Phi}_n}}=C \left(\dot{\Phi}_n-\dot{\Phi}_m \right)
\end{eqnarray}
which is the charge on the capacitor. From the Lagrange equation of motion, we obtain
\begin{eqnarray}
\partial_t Q_n+{ 1 \over L}\left({\Phi}_n-{\Phi}_m \right)=0
\end{eqnarray}
which describes the conservation of the local charge at the node $n$.
The quantization is done imposing the canonical quantization condition
$[\Phi_n,Q_m]= i \hbar \delta_{n m}, \quad [\Phi_n,\Phi_m]=0, \quad [Q_n,Q_m]=0$, 
where $[A,B]$ is the commutator given by $[A,B]=AB-BA$.

The same procedure is applied to the Josephson junction depicted in Fig.~\ref{JJCircuitElements}.
It is composed of two circuit elements, one for electron-tunneling term characterized by the Josephson energy $E_J$, and the other for the capacitor with the capacitance $C_J$.
The Lagrangian for it is given by 
\begin{eqnarray}
{\cal L}_{JJ}={ 1 \over {2}}C_J \left(\dot{\Phi}_n-\dot{\Phi}_m \right)^2+E_J \cos \phi
\end{eqnarray}
where the angular variable $\phi$ satisfies the Josephson relation
\begin{eqnarray}
\dot{\phi}={{2eV} \over \hbar}={{2e} \over \hbar}(\dot{\Phi}_n-\dot{\Phi}_m)
\label{eqJosephson}
\end{eqnarray}
according to Eq.~(\ref{JJdt})
with $V$ being the voltage across the junction. 

It is tempting to use the relation ${\phi}={{2e} \over \hbar} ({\Phi}_n-{\Phi}_m)$
by integrating Eq.~(\ref{eqJosephson}) as has been done in the standard theory.
However, this relation is actually ill-defined since the left-hand side is non-gauge-invariant although the right-hand side is gauge invariant.
Actually, the correct $\phi$ is $\phi^{\rm new}$ in Eq.~(\ref{eq3-new}) according to the new theory. It contains two sources of voltage; the one applicable for the LC circuit case is that arising from ${\bf A}$. Thus, we use the one using ${\bf A}$, and rewrite Eq.~(\ref{eq1}) as
\begin{eqnarray}
\Phi_n=\int_{\rm origin }^{n}{\bf A} \cdot d{\bf r}
\label{Phi-A}
\end{eqnarray}
where the gauge ${\bf E}=-\partial_t {\bf A}$ is adopted. The lower limit of the integration, `origin',  will be fixed for convenience depending on the problem.

The current density ${\bf j}$ is given by
\begin{eqnarray}
{\bf j}=-{{\delta E[{\bf A}]} \over {\delta {\bf A}}}
\end{eqnarray}
in general \cite{Schafroth}, 
where $E[{\bf A}]$ is the total energy expressed as a functional of the vector potential ${\bf A}$. By applying this to the circuit case,
the current through the Josephson junction existing is calculated as 
\begin{eqnarray}
J
={  {\partial {\cal L}_{JJ} } \over {\partial \int_m^n{\bf A} \cdot d{\bf r} } }
=-{ e \over \hbar}E_J \sin \phi^{\rm new}
\end{eqnarray}
The comparison of it with Eq.~(\ref{eqJ1}) gives $J_c=-{ e \over \hbar}E_J$.


\begin{figure}[H]
  \begin{center}
    \begin{circuitikz}[american]
 \draw (0,0)   node[ground]{} node[left]{$1$} 
      to  [josephson=$E_J {,} \, C_J$,-*] (0,2) node[above]{$3$} 
     to[short] (1,2)
   to[josephson=$E_J {,} \, C_J$] (1,0)
      to[short,-*] (0,0);
      \draw (1,2) 
      to[short] (2,2)
        to [C=$C_g$,-*] (4,2) node[above]{$2$} 
      to[V=$V_g$] (4,0)
      to[short] (1,0);
    \end{circuitikz}
  \end{center}
  \caption{A Cooper pair box with two identical Josephson junctions (left junction and right junction) with Josephson energy $E_J$ and capacitance $C_J$. Here, a Josephson junction depicted in Fig.~\ref{JJCircuitElements} is denoted by a single symbol ``$\boxtimes$''. $C_g$ is the capacitance of the gate capacitor, and $V_g$ is the gate voltage provided by the voltage source.}
  \label{CPB2}
\end{figure}
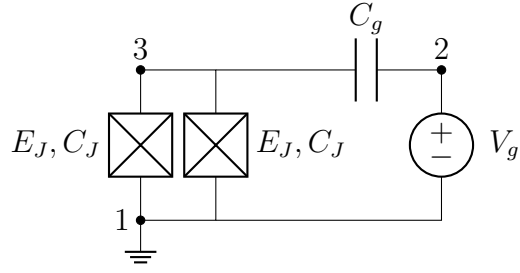

As an example using the dynamical variable in Eq.~(\ref{Phi-A}), we consider the ``Cooper-pair box'' depicted in Fig.~\ref{CPB2}. Its Lagrangian is given by
\begin{eqnarray}
{\cal L}_{CPB}&=&{1 \over {2 }}C_g(\dot{\Phi}_3-\dot{\Phi}_2)^2
\nonumber
\\
&&+ \left[ {1 \over {2 }}C_J(\dot{\Phi}_3-\dot{\Phi}_1)^2
+E_J \cos 
\left( {1 \over 2} (\chi_1 - \chi_3)
+
{ {e} \over {\hbar } }( {\Phi}_3-{\Phi}_1) 
\right) \right]_{\rm L}
\nonumber
\\
&&+ \left[ {1 \over {2 }}C_J(\dot{\Phi}_3-\dot{\Phi}_1)^2
+E_J \cos 
\left( {1 \over 2} (\chi_1 - \chi_3)
+
{ {e} \over {\hbar } }( {\Phi}_3-{\Phi}_1) 
\right) \right]_{\rm R}
\nonumber
\\
\end{eqnarray}
where $[ \cdots ]_{\rm R}$ and $[ \cdots ]_{\rm L}$ denote contributions from the left Josephson junction and the right one, respectively.
Note that the new dynamical variables can handle the path dependence of the values; on the other hand, the original dynamical variables in Eq.~(\ref{eq1}) cannot.

Using a trigonometric identity, it is rewritten as
\begin{eqnarray}
{\cal L}_{CPB}&=&{1 \over {2 }}C_g(\dot{\Phi}_3-\dot{\Phi}_2)^2+{1 \over {2 }}C_J(\dot{\Phi}_3-\dot{\Phi}_1)_{\rm L}^2 +{1 \over {2 }}C_J(\dot{\Phi}_3-\dot{\Phi}_1)_{\rm R}^2
\nonumber
\\
&&+2 E_J 
\cos 
\left( {1 \over 4} [(\chi_1 - \chi_3)_{\rm L}+(\chi_1 - \chi_3)_{\rm R}]
+{ {e} \over {2 \hbar } }[( {\Phi}_3-{\Phi}_1)_{\rm L}+( {\Phi}_3-{\Phi}_1)_{\rm R}] \right)
\nonumber
\\
&& \times 
\cos 
\left( {1 \over 4} [(\chi_1 - \chi_3)_{\rm L}-(\chi_1 - \chi_3)_{\rm R}]
+{ {e} \over {2 \hbar } }[( {\Phi}_3-{\Phi}_1)_{\rm L}- ( {\Phi}_3-{\Phi}_1)_{\rm R}] \right)
\nonumber
\\
\end{eqnarray}
The charge on the node $3$, $Q_3$, is given by
\begin{eqnarray}
Q_3={ {\partial {\cal L}_{CPB} } \over {\partial \dot{\Phi}_3} }=C_g(\dot{\Phi}_3-\dot{\Phi}_2)+C_J(\dot{\Phi}_3-\dot{\Phi}_1)_{\rm L}+C_J(\dot{\Phi}_3-\dot{\Phi}_1)_{\rm R}
\end{eqnarray}
The Lagrange equation for $\Phi_3$ and $\dot{\Phi}_3$ gives rise to the equation for the conservation of the charge on the node $3$,
\begin{eqnarray}
\partial_t{Q_3}+
{{e E_J^{\rm eff}} \over \hbar}  \sin
\left( {\pi \over 2} w_{JJ}[\chi]+{1 \over 2}(\chi_1 - \chi_3)_{\rm R}
+{ {e} \over {2 \hbar } }\Phi_{JJ}+{ {e} \over { \hbar } }( {\Phi}_3-{\Phi}_1)_{\rm R}
\right)
=0
\nonumber
\\
\label{conservationQ}
\end{eqnarray}
where $w_{JJ}[\chi]$ is the winding number for $\chi$ around the loop ${\cal C}_{JJ}$ constructed by the two Josephson junctions,
\begin{eqnarray}
w_{JJ}[\chi] = { 1 \over {2\pi}}[(\chi_1 - \chi_3)_{\rm L}-(\chi_1 - \chi_3)_{\rm R}]
={ 1 \over {2\pi}}\oint_{{\cal C}_{JJ}} \nabla \chi \cdot d {\bf r}
\label{conservationW}
\end{eqnarray}
The value of it is specified as an external parameter.
Another parameter $\Phi_{JJ}$ is the magnetic flux enclosed by ${\cal C}_{JJ}$, 
\begin{eqnarray}
\Phi_{JJ} = (\Phi_1 - \Phi_3)_{\rm L}-(\Phi_1 - \Phi_3)_{\rm R}
=\oint_{{\cal C}_{JJ}} {\bf A} \cdot d {\bf r}
\end{eqnarray}
and
$E_J^{\rm eff}$ is the effective junction energy for the two junctions considered as one,
\begin{eqnarray}
E_J^{\rm eff}=2E_J \cos 
\left({\pi \over 2} w_{JJ}[\chi]+ { {e} \over {2 \hbar } }\Phi_{JJ} \right)
\end{eqnarray}
This depends on $\Phi_{JJ}$ and $w_{JJ}[\chi]$,  may be used to tune the Josephson junction.

We take the node $1$ as `origin' in Eq.~(\ref{Phi-A}); thus, we have $\Phi_1 =0$. We also have $(\dot{\Phi}_1-\dot{\Phi}_2)_{\rm L}=(\dot{\Phi}_1-\dot{\Phi}_2)_{\rm R}=V_g$ 
since there is a voltage generator between the nodes $1$ and $2$.
Overall, the Lagrangian is given by
\begin{eqnarray}
{\cal L}_{CPB}&=&{1 \over {2}}(C_g+2C_J)\dot{\Phi}_3^2+ C_g V_g \dot{\Phi}_3+{1 \over {2}}C_g\dot{\Phi}_3^2
\nonumber
\\
&&+E_J^{\rm eff}\cos \left( {\pi \over 2} w_{JJ}[\chi]+{ {e} \over {2 \hbar } }\Phi_{JJ}+{1 \over 2}(\chi_1 - \chi_3)_{\rm R}
+{ {e} \over { \hbar } }({\Phi}_3)_{\rm R}
\right)
\end{eqnarray}
with
$
Q_3=(C_g+2C_J)\dot{\Phi}_3+V_g C_g 
$. Note that $(\Phi_3)_{\rm R}$ and $(\Phi_3)_{\rm L}$ may differ by a constant, $\Phi_{JJ}$, however, their time-derivatives are the same, which is denoted by $\dot{\Phi}_3$.

Neglecting a constant term, the Hamiltonian is obtained from ${\cal L}_{CPB}$ as
\begin{eqnarray}
{\cal H}_{CPB}&=&{1 \over {2(C_g+2C_J)}} (Q_3-C_g V_g)^2
\nonumber
\\
&&-E_J^{\rm eff}\cos \left( {\pi \over 2} w_{JJ}[\chi]+{ {e} \over {2 \hbar } }\Phi_{JJ}+{1 \over 2}(\chi_1 - \chi_3)_{\rm R}
+{ {e} \over { \hbar } }( {\Phi}_3)_{\rm R}
\right)
\end{eqnarray}
We introduce $\hat{n}$ and $\hat{x}$ by
\begin{eqnarray}
\hat{n}={Q_3 \over {-e}}, \quad {{\hat{x}}}= { {2e} \over {\hbar } }({\Phi}_3)_{\rm R}
\label{electron-number}
\end{eqnarray}
The canonical quantization condition
\begin{eqnarray}
[Q_3, (\Phi_3)_{\rm R}]=-i\hbar
\end{eqnarray}
yields
\begin{eqnarray}
[\hat{n}, {\hat{x} \over 2}]=i
\label{eq32}
\end{eqnarray}
The quantized Hamiltonian is given by
\begin{eqnarray}
{\cal H}_{CPB}=E_C (\hat{n}-n_g)^2
-E_J^{\rm eff}\cos \left( P_{JJ}
+ {\hat{x} \over 2}
\right)
\label{eq58}
\end{eqnarray}
where
\begin{eqnarray}
E_C &=& {e^2 \over {2(C_g+2C_J)}} , \quad n_g={{C_g V_g} \over {e}},
\nonumber
\\
P_{JJ}&=&{\pi \over 2} w_{JJ}[\chi]+{ {e} \over {2 \hbar } }\Phi_{JJ}+{1 \over 2}(\chi_1 - \chi_3)_{\rm R}
\end{eqnarray}
From the relation in Eq.~(\ref{eq32}), the following commutation relations are obtained
\begin{eqnarray}
[e^{\pm i {\hat{x} \over 2}}, \hat{n}]=\pm e^{\pm i {\hat{x} \over 2}}
\label{NumberChange}
\end{eqnarray}
This indicates that $e^{\pm i {\hat{x} \over 2}}$ is the number-changing operator.
Here, the number is that for electrons as indicated in Eq.~(\ref{electron-number}).
We introduce the eigenstate of $\hat{n}$, $|n \rangle$, that satisfies
\begin{eqnarray}
\hat{n} | n \rangle =n| n \rangle
\end{eqnarray}
where $n$ is an integer.
Then, we have the relation
\begin{eqnarray}
e^{\pm i {\hat{x} \over 2}}| n \rangle =| n \pm 1 \rangle
\end{eqnarray}
from Eq.~(\ref{NumberChange}).
Finally, ${\cal H}_{CPB}$ is expressed using the basis $\{ | n \rangle \}$ as
\begin{eqnarray}
{\cal H}_{CPB}=\sum_n 
\left\{ E_C (n-n_g)^2 | n \rangle \langle n|
-{E_J^{\rm eff} \over 2} \left[ e^{i P_{JJ}}
 | n+1 \rangle \langle n|+e^{-i P_{JJ}}| n \rangle \langle n+1| \right] \right\}
\nonumber
\\
\label{eq36}
\end{eqnarray}
The tunneling through the junction in this Hamiltonian is the single-electron one.

In general, if we treat the electronic system of the superconducting electrode denoted by the node $3$ as an isolated system, odd $n$ $|n \rangle$ states has an energy larger than the even one, roughly, by the energy gap $\Delta$ due to the pairing energy gap generation.
Thus, if we include this even-odd energy difference, the Hamiltonian becomes
\begin{eqnarray}
{\cal H}_{CPB}^{\rm new}={\cal H}_{CPB}+ \sum_{n \in {\rm odd }} \Delta| n \rangle \langle n|
\end{eqnarray}
If the value of $\Delta$ is sufficiently large, the ground state will be mostly composed of even $n$ states.
As seen in Fig.~\ref{CPB2}, the node $1$ is susceptible to the noisy environment;
the noise caused on it will affect the electronic state on the node $3$ through 
the single-electron tunneling.  This may explain the quasiparticle poisoning.

When the winding number $w_{JJ}[\chi]$ is zero, $E_J^{\rm eff}$ is given by $2E_J \cos 
\left({ {e} \over {2 \hbar } }\Phi_{JJ} \right)$, thus, the effective junction energy can be modified by the flux $\Phi_{JJ}$ as is usually done in real Josephson junction qubits.
Since the phase factors $e^{\pm i P_{JJ}}$ can be removed by redefining the basis $\{ |n \rangle \} \rightarrow \{ e^{i nP_{JJ}}|n \rangle \}$, eigenvalues are independent of $P_{JJ}$.

Values of $(\chi_1 - \chi_3)_{\rm L}$ and $(\chi_1 - \chi_3)_{\rm R}$ are obtained from relations
in Eqs.~(\ref{conservationQ}) and (\ref{conservationW}) with treating $\Phi_{JJ}$ and $w_{JJ}[\chi]$ as given parameters, and replacing $\partial_t{Q_3}$ and $({\Phi}_3-{\Phi}_1)_{\rm R}$ by their expectation values. This procedure has been employed in the application of the new theory
to some superconducting systems \cite{koizumi2022b}.

\section{The Field operator of electrons for superconducting states and single electron supercurrent tunneling across Josephson junction}
\label{sec3}

$\phi^{\rm new}$ in Eq.~(\ref{eq3-new}) is obtained if the Bogoliubov-de Genne formalism of the BCS theory \cite{Bogoliubov58,deGennes} is improved in such a way that the conservation of  the particle number is achieved.
This is achieved using the Berry connection from many-body wave functions \cite{koizumi2019,koizumi2022b}. We will describe it below.

We start with the Hamiltonian of the BCS theory given by
\begin{eqnarray}
H_{\rm BCS}=\sum_{{\bf k}, \sigma} (\epsilon_{\bf k} -E_F)c^{\dagger}_{{\bf k} \sigma}c_{{\bf k} \sigma}
+\sum_{{\bf k}, {\bf l}}V_{{\bf k} {\bf l}}c^{\dagger}_{{\bf k} \uparrow}c^{\dagger}_{-{\bf k} \downarrow}c_{-{\bf l}  \downarrow}c_{{\bf l} \uparrow}
\label{H-BCS}
\end{eqnarray}
where $c_{{\bf k} \sigma}^\dagger$ and $c_{{\bf k} \sigma}$ are creation and  annihilation operators for the conduction electron with effective mass $m^{\ast}$, the wave vector ${\bf k}$, and spin $\sigma$, respectively; the normal state single-particle energy is given by $\epsilon_{\bf k} ={{\hbar^2 k^2 }\over {2 m^{\ast}}}$, and $E_F$ is the Fermi energy; $V_{{\bf k} {\bf l}}$ is the parameter for effective electron-electron interaction.
Note that associated with $H_{\rm BCS}$ is the field operator
\begin{eqnarray}
\hat{\Psi}_{\sigma}({\bf r})
=
\sum_{i} {1 \over \sqrt{\cal V}}e^{i {\bf k} \cdot {\bf r}} c_{{\bf k} \sigma }
\label{f-BCS}
\end{eqnarray}
where ${\cal V}$ is the volume of the system. 
The physical meaning of this field operator will be the following: The creation of an electron with spin $\sigma$ at the coordinate ${\bf r}$ is expressed by $\hat{\Psi}_{\sigma}^\dagger({\bf r})$; i.e., it is achieved by adding an electron on the linear combination of the single particle state. In the following this field operator is expressed in several different manners,
with changing physical interpretations.

One reason that makes the BCS theory successful is the 
use of the following variational superconducting ground state
\begin{eqnarray}
|{\rm BCS}\rangle =
\prod_{\bf k}(u_{\bf k}+e^{i\theta}v_{\bf k}c^{\dagger}_{{\bf k} \uparrow} c^{\dagger}_{-{\bf k} \downarrow} )|{\rm vac} \rangle
\label{theta1}
\end{eqnarray}
where $|{\rm vac} \rangle$ is the vacuum state that satisfies 
$c_{{\bf k} \sigma}|{\rm vac} \rangle=0$, and $u_{\bf k}$ and $v_{\bf k}$ are real variational parameters satisfying $u_{\bf k}^2+v_{\bf k}^2=1$ \cite{BCS1957}. A salient feature of this state vector is that expectation values for some operators calculated by it are not zero, although they are usually zero due to the conservation of the particle number \cite{Peierls1991}; 
\begin{eqnarray}
\langle {\rm BCS}| c^{\dagger}_{{\bf k} \uparrow}c^{\dagger}_{-{\bf k} \downarrow}|{\rm BCS}\rangle \neq 0, \quad 
\langle {\rm BCS}|c_{-{\bf l}  \downarrow}c_{{\bf l} \uparrow}|{\rm BCS}\rangle \neq 0
\label{eq0}
\end{eqnarray}
This non-conservation of the particle number in the BCS theory makes the phase $\theta$ in $|{\rm BCS}\rangle$ physically, meaningful. The Josephson effects occur due to this physically meaningful $\theta$.
The energy gap is calculated using Eq.~(\ref{eq0}) as
\begin{eqnarray}
\Delta_{\bf k}=-\sum_{\bf l} V_{{\bf l} {\bf k}} \langle {\rm BCS}|c_{-{\bf l}  \downarrow}c_{{\bf l} \uparrow}|{\rm BCS}\rangle=-e^{i \theta}\sum_{\bf l} V_{{\bf l} {\bf k}}u_{\bf l}v_{\bf l}
\label{E-gap}
\end{eqnarray}
Note that the supercurrent is assuemd to arise from the spatial variation 
of $\theta$ in the standard theory.

Now consider the Josephson junction. It has two superconducting electrodes. We use the following left (denoted by $L$) and right (denoted by $R$) BCS states for them
\begin{eqnarray}
|{\rm BCS}_L\rangle &=&
\prod_{\bf k}(u_{L {\bf k}}+e^{i\theta_L} v_{L {\bf k}}c^{\dagger}_{L {\bf k} \uparrow} c^{\dagger}_{L -{\bf k} \downarrow} )|{\rm vac} \rangle, 
\nonumber
\\
|{\rm BCS}_R\rangle &=&
\prod_{\bf k}(u_{R {\bf k}}+e^{i\theta_R} v_{R {\bf k}}c^{\dagger}_{R {\bf k} \uparrow} c^{\dagger}_{R -{\bf k} \downarrow} )|{\rm vac} \rangle
\nonumber
\\
\label{thetaLR}
\end{eqnarray}
where quantities for the left superconductor is labeled with $L$, and those for the right superconductor are labeled with $R$. The supercurrent arises from the phase difference $\theta_R-\theta_L$.

The hopping Hamiltonian between the two superconductors is given by
\begin{eqnarray}
H_{LR}=-\sum_{\sigma}T_{LR}e^{i {e \over {\hbar } } \int^{R}_{L} d{\bf r} \cdot {\bf A}}
  c^{\dagger}_{L \sigma} c_{R \sigma}+{\rm h.c.}
\label{juncH}
\end{eqnarray}
and it is taken as a perturbation to produce the Josephson effects. Here, $T_{LR}$ is the hoping matrix element of the junction, and  $c^{\dagger}_{L \sigma}$ and  $c_{R \sigma}$ denote
creation and annihilation operators for electrons with spin $\sigma$, respectively. 
The tunneling occurs as the second order perturbation of $H_{LR}$, $H_{LR}{1 \over {E_0 - H_0}} H_{LR}$, as the effective perturbation ($H_0$ is the zeroth order Hamiltonian, and $E_0$ is the ground state energy), and $\phi$ is given by
\begin{eqnarray}
\phi=\theta_R-\theta_L+{{2e}\over {\hbar}}
\int^{R}_{L}
 {\bf A} \cdot d{\bf r}
 \label{eq:phi}
\end{eqnarray}
This corresponds to $\phi^{\rm old}$ in Eq.~(\ref{eq3}).

We may regard the superconducting ground state as a vacuum of the Bogoliubov quasiparticles  \cite{Bogoliubov58}.
For the state vector $|{\rm BCS}\rangle$ case, the Bogoliubov operators are defined by
\begin{eqnarray}
\gamma_{{\bf k} \sigma}=e^{-{i \over 2} \theta}u_{\bf k} c_{{\bf k} \sigma} - \sigma e^{{i \over 2} \theta} v_{\bf k}c^{\dagger}_{-{\bf k} -\sigma} 
\label{BogolubovBCS}
\end{eqnarray}
where $\sigma=1, \sigma =-1$ indicate up-spin and down-spin, respectively.
They are fermion operators, and satisfy
\begin{eqnarray}
\gamma_{{\bf k} \sigma}|{\rm BCS} \rangle =0
\label{eqBog}
\end{eqnarray}
This indicate that the superconducting ground state can be regarded as a vacuum of the Bogoliubov quasiparticles.

Form Eq.~(\ref{BogolubovBCS}), we obtain,
\begin{eqnarray}
c_{{\bf k} \sigma}=e^{{i \over 2}\theta} [u_{\bf k}\gamma_{{\bf k} \sigma}+\sigma v_{\bf k}\gamma^\dagger_{-{\bf k} -\sigma}]
\end{eqnarray}
where the fact is used that real parameters $u_{\bf k}$ and $v_{\bf k}$ satisfy the relations $u_{\bf k}=u_{-{\bf k}}$, $v_{\bf k}=v_{-{\bf k}}$, 
and $u^2_{\bf k}+v^2_{{\bf k}}=1$. Then, the field operator is expressed as
\begin{eqnarray}
\hat{\Psi}_{\sigma}({\bf r})
=
\sum_{i} {1 \over \sqrt{\cal V}}e^{i {\bf k} \cdot {\bf r}} 
e^{{i \over 2}\theta} [u_{\bf k}\gamma_{{\bf k} \sigma}+\sigma v_{-{\bf k}}\gamma^\dagger_{-{\bf k} -\sigma}
]
\label{f-BCS-Bog}
\end{eqnarray}
The creation of an electron with spin $\sigma$ at the coordinate ${\bf r}$ expressed by $\hat{\Psi}_{\sigma}^\dagger({\bf r})$ is now achieved by the superposition of adding an electron and removing it since the
Bogoliubov operators are mixture of the creation and annihilation operators of electrons.
This sounds awkward since the creation of an electron can be achieved by the annihilation of it.
This awkwardness is removed if the Bogoliubov operators are so modified that they conserve particle numbers, which is done in the new theory.

The new phase $\phi^{\rm new}$ in Eq.~(\ref{eq3-new}) 
arises if we employ new Bogoliubov operators that conserve the particle number.
Let us define $\chi$ appearing in $\phi^{\rm new}$.
The variable $\chi$ is obtained from the Berry connection from many-body wave functions \cite{koizumi2019}, ${\bf A}_{\Psi}^{\rm MB}$, defined by
\begin{eqnarray}
\! \!
\!{\bf A}^{\rm MB}_{\Psi}({\bf r},t)\!=\!
{{{\rm Re} \left\{
 \int d\sigma_1  d{\bf x}_{2}  \cdots d{\bf x}_{N}
 \Psi^{\ast}({\bf r}, \sigma_1, \cdots, {\bf x}_{N},t)
  (-i \hbar \nabla )
\Psi({\bf r}, \sigma_1, \cdots, {\bf x}_{N},t) \right\}
 }
 \over {\hbar \rho({\bf r},t)}} 
\nonumber
\\
\label{Afic}
\end{eqnarray}
Here,
`$\rm{Re}$' denotes the real part, $\Psi$ is the total electronic wave function, ${\bf x}_i$ collectively stands for the coordinate ${\bf r}_i$ and the spin $\sigma_i$ of the $i$th electron, $-i \hbar \nabla$ is the Schr\"{o}dinger's momentum operator for the coordinate vector ${\bf r}$, and $\rho({\bf r},t)$ is the number
density calculated from $\Psi$. This Berry connection is obtained by regarding ${\bf r}$ as the ``adiabatic parameter''\cite{Berry}. 

The variable $\chi$ is defined from the above Berry connection as 
 \begin{eqnarray}
{ {\chi({\bf r},t)}}= - 2\int^{{\bf r}}_0 {\bf A}_{\Psi}^{\rm MB}({\bf r}',t) \cdot d{\bf r}' 
\end{eqnarray}
and the total electronic wave function
 $\Psi$ is given by
\begin{eqnarray}
\Psi({\bf x}_1, \cdots, {\bf x}_N,t)=\exp \left(  -{i \over 2} \sum_{j=1}^{N} \chi({\bf r}_j,t)\right)
\Psi_0({\bf x}_1, \cdots, {\bf x}_N,t)
\label{single-valued}
\end{eqnarray}
with $\Psi_0({\bf x}_1, \cdots, {\bf x}_N,t)$ denoting the currentless wave function. It is important to note the current is all due to the phase factor $\exp \left(  -{i \over 2} \sum_{j=1}^{N} \chi({\bf r}_j,t)\right)$.
The wave function $\Psi$ corresponds to the case where all the electrons are participating in the collective mode $\chi$. We denote the corresponding state vector as $|{\rm Cnd}(N) \rangle$.

Now we go beyond the Schr\"{o}dinger representation of quantum mechanics by takeing into account the fluctuation of the number of electrons participating in the collective $\chi$ mode. Then, the electronic state cannot be described by the wave function in Eq.~(\ref{single-valued}).
The $\chi$ mode is the whole system motion described by $\exp \left(  -{i \over 2} \sum_{j=1}^{N} \chi({\bf r}_j,t)\right)$ for the case where the number of particles participating in it is $N$. We take into account of the case where
 this number varies in the rage $0,1,\cdots, N$.
  For this purpose, we can quantize the $\chi$ mode as bosons. The quantization procedure is explained, previously \cite{koizumi2022b}. 
  The number changing operators $e^{{i \over 2}\hat{X}}$ and $e^{-{i \over 2}\hat{X}}$ given below are obtained from this quantization.
The operator $e^{-i \hat{X}}$ is an operator that reduces the number of electrons participating in the $\chi$ mode by two given by
\begin{eqnarray}
e^{-i \hat{X}}=|{\rm Cnd}(0) \rangle \langle {\rm Cnd}(2)|+ \cdots + |{\rm Cnd}(N-2) \rangle \langle {\rm Cnd}(N)|
\label{chi-1}
\end{eqnarray}
and  $e^{i \hat{X}}$increases by two, given by
\begin{eqnarray}
e^{i \hat{X}}=|{\rm Cnd}(2) \rangle \langle {\rm Cnd}(0)|+ \cdots + |{\rm Cnd}(N) \rangle \langle {\rm Cnd}(N-2)|
\label{chi-2}
\end{eqnarray}

In the new formalism, the superconducting state is expressed as
\begin{eqnarray}
|{\rm Gnd}(N)\rangle =
\prod_{\bf k}(u_{\bf k}+v_{\bf k} c^{\dagger}_{{\bf k} \uparrow} c^{\dagger}_{-{\bf k} \downarrow}e^{-i \hat{X}} )|{\rm Cnd}(N) \rangle
\label{chi1}
\end{eqnarray}
 Since 
$e^{-i \hat{X}}$ reduces the electron number by two and $c^{\dagger}_{{\bf k} \uparrow} c^{\dagger}_{-{\bf k}}$ increases by two, the particle number is fixed in $|{\rm Gnd}(N)\rangle$. 

Using $e^{\pm i \hat{X}}$, $H_{\rm BCS}$ in Eq.~(\ref{H-BCS}) can be modified as
\begin{eqnarray}
\bar{H}_{\rm BCS}=\sum_{{\bf k}, \sigma} (\epsilon_{\bf k} -E_F)c^{\dagger}_{{\bf k} \sigma}c_{{\bf k} \sigma}
+\sum_{{\bf k}, {\bf l}}V_{{\bf k} {\bf l}}c^{\dagger}_{{\bf k} \uparrow}c^{\dagger}_{-{\bf k} \downarrow}e^{-i \hat{X}}e^{i \hat{X}}c_{-{\bf l}  \downarrow}c_{{\bf l} \uparrow}
\label{H-BCS-new}
\end{eqnarray}
Then, the relations in Eq.~(\ref{eq0}) are replaced by the particle-number conserving ones
\begin{eqnarray}
\langle {\rm Gnd}(N)| c^{\dagger}_{{\bf k} \uparrow}c^{\dagger}_{-{\bf k} \downarrow}e^{-i \hat{X}}|{\rm Gnd}(N)\rangle \neq 0, \quad 
\langle {\rm Gnd}(N)|e^{i \hat{X}} c_{-{\bf l}  \downarrow}c_{{\bf l} \uparrow}|{\rm Gnd}(N)\rangle \neq 0
\label{eq0-new}
\end{eqnarray}
and the new energy gap equation is given by
\begin{eqnarray}
-\sum_{\bf l} V_{{\bf l} {\bf k}} \langle {\rm Gnd}(N)|e^{i \hat{X}}c_{-{\bf l}  \downarrow}c_{{\bf l} \uparrow}|{\rm Gnd}(N)\rangle=-\sum_{\bf l} V_{{\bf l} {\bf k}}u_{\bf l}v_{\bf l}
\label{E-gap2}
\end{eqnarray}
This is equal to the result in Eq.~(\ref{E-gap}) with $\theta =0$ \cite{BCS1957}. Actually, the original BCS paper deals with this case.
Equations for $u_{\bf k}$, $v_{\bf k}$, and $\Delta_{\bf k}$ are the same as the BCS theory, thus, major results of the BCS theory are unaltered in the new theory. 

Corresponding to the Bogoliubov operators in Eq.~(\ref{BogolubovBCS}), the new Bogoliubov operators are
\begin{eqnarray}
\bar{\gamma}_{{\bf k} \sigma}=u_{\bf k} e^{{i \over 2} \hat{X}}  c_{{\bf k} \sigma}-\sigma v_{\bf k}  c_{-{\bf k} -\sigma}^{\dagger} e^{-{i \over 2} \hat{X}} 
\label{Bog-new-K}
\end{eqnarray}
A salient feature of them is that they conserve the number of particles since $e^{{i \over 2} \hat{X}}$ increases the particle number by one, and $e^{-{i \over 2} \hat{X}}$ decreases the particle number by one.
They are fermion number operators, and satisfy
\begin{eqnarray}
\bar{\gamma}_{{\bf k} \sigma}|{\rm Gnd}(N)\rangle =0
\label{eqBog-new}
\end{eqnarray}


The electron field operator is now expressed using the new Bogolubov operators and $e^{-{i \over 2} \hat{X}}$ as
\begin{eqnarray}
\hat{\Psi}_{\sigma}({\bf r})
=
\sum_{i} {1 \over \sqrt{\cal V}}e^{i {\bf k} \cdot {\bf r}} e^{-{i \over 2} \hat{X}}
[u_{\bf k} \bar{\gamma}_{{\bf k} \sigma}+\sigma v_{\bf k} \bar{\gamma}^\dagger_{-{\bf k} -\sigma}]
\end{eqnarray}
This indicates that the creation of an electron with spin $\sigma$ at the coordinate ${\bf r}$ is achieved 
by first adding it to the $\chi$ mode by the operation of $e^{{i \over 2} \hat{X}}$, and then, transferred to the single particle state by the 
operation of $\sum_{i} {1 \over \sqrt{\cal V}}e^{-i {\bf k} \cdot {\bf r}}
[u_{\bf k} \bar{\gamma}_{{\bf k} \sigma}^\dagger +\sigma v_{\bf k} \bar{\gamma}_{-{\bf k} -\sigma}]$. The original 
$H_{\rm BCS}$ cannot include such effects, however, the modified one, $\bar{H}_{\rm BCS}$, in Eq.~(\ref{H-BCS-new}) can.

In the standard theory based on the BCS one, the supercurrent is generated by the spatial variation of $\theta$. In the new theory, it is due to the spatial variation of $\chi$.
In order to include the latter, we utilize the generalized Bogoliubov operator formalism using the Wannier basis \cite{deGennes,Zhu2016}.
First, the BCS case is examined; it uses the electron field operator given by
\begin{eqnarray}
\hat{\Psi}_{\sigma}({\bf r})
=
\sum_{i} w({\bf r}-{\bf r}_i) c_{i \sigma }
\label{f-original}
\end{eqnarray}
where $w({\bf r}-{\bf r}_i) c_{i \sigma }$ is a localized (Wannier) orbital around ${\bf r}={\bf r}_i$, and $c_{i \sigma }$ is the electron annihilation operator for this orbital state with spin $\sigma$. 
Next, transform the above to the one expressed by the Bogoliubov operators ${\gamma}^\dagger_{{n} \sigma }$ and ${\gamma}_{{n} \sigma }$
\begin{eqnarray}
\hat{\Psi}_{\sigma}({\bf r})
=
\sum_{n} \left[ \gamma_{{n} \sigma} u_{n}({\bf r}) -\sigma \gamma_{{n} \, -\sigma }^{\dagger} v^{\ast}_{n}({\bf r}) \right]
\label{f2}
\end{eqnarray}
where  ${\gamma}_{{n} \sigma }$ satisfy 
\begin{eqnarray}
\gamma_{n \sigma}|{\rm BCS}\rangle=0
\end{eqnarray}
and $u_{\bf k}, v_{\bf k}$ in Eq.~(\ref{BogolubovBCS}) are replaced by the coordinate dependent ones, $u_{n}({\bf r}), v_{n}({\bf r})$ \cite{deGennes}.

Let us move to the particle number conserving  Bogoliubov operator case.
The electron field operator in this case is expressed as 
\begin{eqnarray}
\hat{\Psi}_{\sigma}({\bf r},t)&=&\sum_{n} e^{-{i \over 2}\hat{\chi} ({\bf r},t)}\left[ \bar{\gamma}_{{n} \sigma } u_{n}({\bf r})  -\sigma \bar{\gamma}^{\dagger}_{{n} \, -\sigma } v^{\ast}_{n}({\bf r}) \right]
\label{f2-new}
\end{eqnarray}
where the coordinate dependent new Bogoliubov operators $\bar{\gamma}_{{n} \sigma }$ satisfy
\begin{eqnarray}
\bar{\gamma}_{n \sigma}|{\rm Gnd}(N) \rangle=0
\end{eqnarray}
The coordinate dependent version of $\hat{X}$ is denoted by $\hat{\chi}({\bf r},t)$, and satisfies
\begin{eqnarray}
e^{\pm {i \over 2} \hat{\chi}({\bf r},t)}|{\rm Cnd}(N) \rangle= e^{\pm {i \over 2} {\chi}({\bf r},t)}|{\rm Cnd}(N\pm1) \rangle
\label{eqBC}
\end{eqnarray}

For a lattice system in which the site $\ell$ corresponds to the coordinate ${\bf r}_\ell$, and the Wannier function is replaced as $w({\bf r}_j-{\bf r}_\ell) \rightarrow \delta_{j \ell}$,
the relation in Eq.~(\ref{f2-new}) indicates that annihilation and creation operators for the electrons at the $i$th site are given by
 \begin{eqnarray}
 c_{ \ell \sigma} &=& e^{ -{i \over 2} \hat{\chi}_\ell} \sum_{n}[ u^{n}_{\ell}\bar{\gamma}_{n \sigma}-\sigma (v^{n}_{\ell})^{\ast}\bar{\gamma}_{n -\sigma}^{\dagger}] 
 \nonumber
 \\
 c^{\dagger}_{\ell \sigma} &=&  \sum_{n} [ (u^{n}_{\ell \sigma})^{\ast}\bar{\gamma}^{\dagger}_{n \sigma}-\sigma v^{n}_{\ell}\bar{\gamma}_{n -\sigma}] e^{{i \over 2} \hat{\chi}_\ell}
 \label{Bog}
  \end{eqnarray}
where $u_\ell^n$, $v_\ell^n$, and $\hat{\chi}_\ell$ stand for $u_n({\bf r}_\ell)$, $v_n({\bf r}_\ell)$, and $\hat{\chi}({\bf r}_\ell, t)$, respectively. 
In the new theory, the creation of a single electron at the $\ell$th site with spin $\sigma$ by $c^{\dagger}_{\ell \sigma}$ becomes a successive processes of the creation of an electron in the collective mode at the $\ell$th site by $e^{{i \over 2} \hat{\chi}_\ell}$, 
and subsequent changing of it to the single electron state at the $\ell$th site with spin $\sigma$ by the application of the operator $\sum_{n} [ (u^{n}_{\ell \sigma})^{\ast}\bar{\gamma}^{\dagger}_{n \sigma}-\sigma v^{n}_{\ell}\bar{\gamma}_{n -\sigma}]= c^{\dagger}_{\ell \sigma} e^{-{i \over 2} \hat{\chi}_\ell} $. 
A method to solve problems on superconductors using the field operator in Eq.~(\ref{f2-new}) and the particle-number conserving  Bogoliubov operators in Eq.~(\ref{Bog}) is explained in Ref.~\cite{koizumi2020c,koizumi2021,Koizumi2021c}. Actually, $\chi({\bf r},t)$ is obtained from the conservation of the local charge (such as Eq.~(\ref{conservationQ})) and
the conservation of the winding number (such as Eq.~(\ref{conservationW})).
When the solution yields a spatially varying $\chi$, the system is equipped with a flow proportional to $\nabla \chi$. This generate supercurrent.  The creation of an electron is affected by it.

Using the creation and annihilation operators in Eq.~(\ref{Bog}), $H_{LR}$ is now expressed as
\begin{eqnarray}
H_{LR}&=&-T_{L R} e^{{ i \over 2}(\hat{\chi}_L-\hat{\chi}_R)} e^{-i {e \over \hbar } \int_{R}^{L} d{\bf r} \cdot {\bf A}}
\times
\sum_{n,m} \Big[
(
(u^{n}_{L})^{\ast}\bar{\gamma}^{\dagger }_{n \downarrow} + v^{n}_{L }\bar{\gamma}_{n \uparrow}) ( u^{m}_{R }\bar{\gamma}_{m \downarrow}+ (v^{m}_{R})^{\ast}\bar{\gamma}_{m \uparrow}^{\dagger} ) 
\nonumber
\\
&&+
(
(u^{n}_{L })^{\ast}\bar{\gamma}^{\dagger }_{n \uparrow} - v^{n}_{L}\bar{\gamma}_{n \downarrow}) ( u^{m}_{R }\bar{\gamma}_{m \uparrow}- (v^{m}_{R})^{\ast}\bar{\gamma}_{m  \downarrow}^{\dagger} )  
\Big]+\mbox{h.c.}
\label{eq39}
\end{eqnarray}
where the new Bogoliubov operators with subscript $\ell=L$ indicates it is for  the left superconductor ($\ell=R$ for the right superconductor).
The single-electron tunneling occurrs as the first order perturbation of Eq.~(\ref{eq39}) due to the fact that the same Bogoliubov operators, $\bar{\gamma}^{\dagger }_{n \sigma}$ and  $\bar{\gamma}_{n \sigma}$, are multiplied by parameters for both left and right superconductors. The phase difference in the Josephson formula $\phi$ in Eq.~(\ref{eqJ1}) can be deduced from the phase factor $e^{{ i \over 2}(\hat{\chi}_L-\hat{\chi}_R)} e^{-i {e \over \hbar } \int_{R}^{L} d{\bf r} \cdot {\bf A}}$ in $H_{LR}$ as
\begin{eqnarray}
\phi=-{1 \over 2}(\chi_R-\chi_L)+{{e}\over {\hbar}}
\int^{R}_{L}
 {\bf A} \cdot d{\bf r}
 \label{eq:phi-new}
\end{eqnarray}
where operators are replaced by their expectation values. This agrees with $\phi^{\rm new}$ in Eq.~(\ref{eq3-new}).

\section{Possible way to avoid `quasiparticle poisoning'}
\label{sec4}
Superconductors usually exhibits stabilization due to the electron pairing. This effect arises from the fact that the pairing creates an energy gap.
In the BCS theory, the supercurrent is the flow of the paired-electrons, and
the single-electron supercurrent tunneling is prohibited.
However, in the new theory, the supercurrent is a flow generated by the collective $\chi$ mode, and the single electron tunneling  is possible. It will spoil the stabilization by the electron-pairing. It may be the cause of the `quasiparticle poisoning' observed in the superconducting qubits with Josephson junctions.
The new theory suggests that this will be avoided if 
the situation is realized where the single-electron supercurrent tunneling is prohibited.
Such a situation  will be achieved if the interaction between the two superconductors in the Josephson junction is so weak that the Bogoliubov operators in the left and right superconductors are different.
 Actually, the standard theory deals with this case.

Let us consider the situation where the Bogoliubov operators in the left and right superconductors are different.
We denote the Bogoliubov operators for the left superconductor as $\bar{\gamma}^{\dagger}_{L n \sigma}$, and 
for the right superconductor as $\bar{\gamma}_{L n \sigma}$, and $\bar{\gamma}^{\dagger}_{R n \sigma}$ and $\bar{\gamma}_{R n \sigma}$.  Then, the electron creation and annihilation operators in the left and right superconductors are expressed as
 \begin{eqnarray}
 c_{ L \sigma} &=&\sum_{n}e^{ -{i \over 2} \hat{\chi}_L}[ u^{n}_{L}\bar{\gamma}_{L n \sigma}-\sigma (v^{n}_{L})^{\ast}\bar{\gamma}_{L n -\sigma}^{\dagger}] 
 \nonumber
 \\
 c^{\dagger}_{ L \sigma} &=&\sum_{n} e^{{i \over 2} \hat{\chi}_L}[ (u^{n}_{L \sigma})^{\ast}\bar{\gamma}^{\dagger}_{L n \sigma}-\sigma v^{n}_{L}\bar{\gamma}_{L n -\sigma}] 
  \nonumber
 \\
  c_{ R \sigma} &=&\sum_{n}e^{ -{i \over 2} \hat{\chi}_R}[ u^{n}_{R}\bar{\gamma}_{R n \sigma}-\sigma (v^{n}_{R})^{\ast}\bar{\gamma}_{R n -\sigma}^{\dagger}] 
 \nonumber
 \\
 c^{\dagger}_{ R \sigma} &=&\sum_{n} e^{{i \over 2} \hat{\chi}_R}[ (u^{n}_{R \sigma})^{\ast}\bar{\gamma}^{\dagger}_{R n \sigma}-\sigma v^{n}_{R}\bar{\gamma}_{R n -\sigma}] 
 \label{BogLR}
  \end{eqnarray}
Then, $H_{LR}$ becomes
  \begin{eqnarray}
H_{LR}&=&-T_{L R} e^{ { i \over 2}(\hat{\chi}_L-\hat{\chi}_R)} e^{-i {e \over {\hbar }} \int_{R}^{L}  d{\bf r} \cdot {\bf A}}
\times
\sum_{n,m} \Big[
(
(u^{n}_{L})^{\ast}\bar{\gamma}^{\dagger }_{L n \downarrow} + v^{n}_{L }\bar{\gamma}_{L n \uparrow}) ( u^{m}_{R }\bar{\gamma}_{R m \downarrow}+ (v^{m}_{R})^{\ast}\bar{\gamma}_{R m \uparrow}^{\dagger} ) 
\nonumber
\\
&&+
(
(u^{n}_{L })^{\ast}\bar{\gamma}^{\dagger }_{L n \uparrow} - v^{n}_{L}\bar{\gamma}_{L n \downarrow}) ( u^{m}_{R }\bar{\gamma}_{R m \uparrow}- (v^{m}_{R})^{\ast}\bar{\gamma}_{R m  \downarrow}^{\dagger} )  
\Big]+\mbox{h.c.}
\end{eqnarray}
In this case, the tunneling does not occur by the first order perturbation of it.

The tunneling occurs by the second order perturbation.
The second order effective Hamiltonian with taking expectation values for the Bogoliubov operators is given by
\begin{eqnarray}
&&\left\langle H_{LR}{1 \over {E_0 - H_0}} H_{LR} \right\rangle
\nonumber
\\
& \approx& - \Big\langle \sum_{m, n, m', n'}T_{L R}^2 
\left[ e^{ -{i \over 2} (\hat{\chi}_L-\hat{\chi}_R)}e^{-i {e \over {\hbar }} \int_{R}^{L}  d{\bf r} \cdot {\bf A}}v_L^{n}u_R^{m}(\bar{\gamma}_{L n \uparrow} \bar{\gamma}_{R m \downarrow}-\bar{\gamma}_{L n \downarrow}\bar{\gamma}_{R m \uparrow})+ (L \leftrightarrow R) 
\right]
\nonumber
\\
&\times& {1 \over {\epsilon_m^{R}+\epsilon_n^{L}}}
\left[ e^{ -{i \over 2} (\hat{\chi}_L-\hat{\chi}_R)}e^{-i {e \over {\hbar }} \int_{R}^{L}  d{\bf r} \cdot {\bf A}}(u_{L}^{ n'}v_{R}^{ m'})^{\ast} (\bar{\gamma}^{\dagger}_{L n' \downarrow} \bar{\gamma}^{\dagger}_{R m' \uparrow}- \bar{\gamma}^{\dagger}_{L n' \uparrow} \bar{\gamma}^{\dagger}_{R m' \downarrow})+ (L \leftrightarrow R) 
\right] \Big\rangle
\nonumber
\\
&\approx& - \sum_{m, n} {{2T_{L R}^2 } \over {\epsilon_m^{R}+\epsilon_n^{L}}}
\Big[ v_L^{n}u_R^{m} (u_{L}^{ n}v_{R}^{ m})^{\ast}e^{- {i } ({\chi}_L-{\chi}_R)-i {{2e} \over {\hbar }} \int_{R}^{L}  d{\bf r} \cdot {\bf A}}
+(v_L^{n}u_R^{m})^{\ast} u_{L}^{ n}v_{R}^{ m}e^{ {i} ({\chi}_L-{\chi}_R)+i {{2e} \over {\hbar }} \int_{R}^{L} d{\bf r} \cdot {\bf A}}
\nonumber
\\
&&+|u_{L}^{ n}v_{R}^{ m}|^2+ |v_{L}^{ n}u_{R}^{ m}|^2\Big]
\label{Perttransfer3}
\end{eqnarray}
where $\langle \hat{O} \rangle$ indicates the ground state expectation value of the operator $\hat{O}$. As the phase factors $e^{\pm {i } [({\chi}_L-{\chi}_R)+ {{2e} \over {\hbar }}\int_{R}^{L} d{\bf r} \cdot {\bf A}]}$ indicate,
the phase $\phi$ in the Josephson current is $\phi^{\rm old}$; the tunneling occurs as the electron-pair tunneling as in the standard theory.
Actually, the Ambegaokar-Baratoff relation \cite{Ambegaokar} is obtained from the above. 

The second order effective hopping Hamiltonian in Eq.~(\ref{Perttransfer3}) does not cause the quasiparticle poisoning.
Thus, a way to avid the quasiparticle poisoning is to weaken the interaction between
two superconducting states across the junction so that
the Bogoliubov operaters in the left and right superconductors become different as shown in Eq.~(\ref{BogLR}). This may be achieved by thickening the insulator part of the junction or raising the temperature. 
For such a weak coupling junction, the time variation of $\phi^{\rm old}$ given by Eq.~(\ref{eqJosephsonOld}) may be observed.

\section{Concluding remarks}
\label{sec5}

The original intention to use the particle-number non-conserving ground state $|{\rm BCS} \rangle$ in Eq.~(\ref{H-BCS}) was to facilitate
calculations involving electron-pairing, assuming that the electron-pairing is the cause of superconductivity \cite{BCS1957}.
 At that time, it was believed that superconductivity was a phenomenon explainable by the Schr\"{o}dinger equation. Thus, the use of the particle-number non-conserving formalism was regarded as merely a practical mathematical tool. 
 
 However, the use of the particle-number non-conserving formalism actually brings in the mathematical structure needed to describe the superconducting state which is outside the Schr\"{o}dinger equation formalism. The explanation of superconductivity  requires  a theory equipped with the mathematical structure allowing the relations in  Eqs.~(\ref{eq0}), (\ref{E-gap}),
(\ref{BogolubovBCS}), and (\ref{eqBog}). 
Since it is sensible to consider that superconductivity occurs in particle-number conserving systems, the particle-number non-conserving formalism should be an approximation \cite{Peierls1991}. Recently, it has been shown that the required mathematical structure with conserving the particle-number is achieved if we  take into account the neglected $U(1)$ phase in the Schr\"{o}dinger representation of quantum mechanics \cite{koizumi2022,koizumi2022b}. Through the quantization of the collective mode arising from it, the required relations are realized as Eqs.~(\ref{eq0-new}), (\ref{E-gap2}), (\ref{Bog-new-K}), and (\ref{eqBog-new}). In this new theory, the role of the electron pairing is to stabilize this collective mode by generating the energy gap; however, the origin of superconductivity is not the electron-pairing but the appearance of the collective $\chi$ mode.

Although the BCS theory has been a successful one, serious disagreements with experiments exist: 1) it uses the particle number non-conserving formalism \cite{Peierls1991}; 2) the superconducting carrier mass obtained by the London moment experiment is the free electron mass $m_e$, although the predicted one by the BCS theory is the effective mass $m^\ast$ of the normal state \cite{Hirsch2013b}; 3) the reversible superconducting-normal phase transition in a magnetic field cannot be explained \cite{Hirsch2017}; 4) the dissipative quantum phase transition in a Josephson junction system predicted by the standard theory is absent \cite{PhysRevX2021a}. We may add the quasiparticle poisoning problem to the above list.
All those problems are resolved in the new theory \cite{koizumi2021,koizumi2022b,koizumi2023}.
Since the new theory reproduces major experimental results explained by the BCS theory and also resolves the above mentioned problems, it my be a superseding theory of the BCS one. 

In this new theory, the electron field operator is given by Eq.~(\ref{f2-new}).
The conventional many-body formalism using the electron field operator given in Eq.~(\ref{f-BCS}) misses effects arising from the $U(1)$ phase neglected by Dirac \cite{koizumi2022,koizumi2022b}. Although the BCS's particle-number
non-conserving formalism takes into account the deviation from such a formalism, 
it uses the particle-number non-conserving formalism.
For the true understanding of superconductivity, the particle-number conserving theory with including the neglected $U(1)$ phase will be required.
The improvement of the performance of superconducting qubits may be achieved based on it.

\

{\bf References}

\

\providecommand{\newblock}{}

\end{document}